\documentclass[doublespacing]{elsart}

\usepackage{graphicx}
\usepackage{amssymb}

\begin{document}

\begin{frontmatter}



\title{Nuclear profile dependence of elliptic flow from a parton cascade}


\author{Bin Zhang}

\address{Department of Chemistry and Physics,
Arkansas State University,\\
P.O. Box 419, State University,
Arkansas 72467-0419, USA}

\begin{abstract}
% Text of abstract
The transverse profile dependence of elliptic flow is studied in
a parton cascade model. 
We compare results from the binary scaling profile to results from
the wounded nucleon scaling profile. 
The impact parameter dependence of elliptic 
flow is shown to depend sensitively on the transverse profile of initial 
particles, however, if elliptic flow is plotted as a function of 
the relative multiplicity, the nuclear profile dependence disappears. 
The insensitivity was found previously in a hydrodynamical calculation. 
Our calculations indicate that the insensitivity is also valid 
with additional viscous corrections. In addition, the minimum bias 
differential elliptic flow is demonstrated to be insensitive to the 
nuclear profile of the system.
\end{abstract}

\end{frontmatter}

% main text
%\section{}
%\label{}

Recently, the Relativistic Heavy Ion Collider has produced large amount of 
exciting new data. These new data give us valuable insight into
the hot and dense nuclear matter. One of the important observables
is elliptic flow which reflects the transverse anisotropy of
particle momentum distribution. Elliptic flow has been studied by
many theoretical models, including non-Abelian energy loss models
\cite{gyulassy01a,gyulassy02a}, 
saturation models \cite{krasnitz02a,teaney02a,kovchegov02a}, 
parton recombination models 
\cite{lin02a,molnar03a,lin03a,fries03a,greco03a,greco03b,fries03b,nonaka03a},
hydrodynamical models \cite{teaney01a,kolb00a,kolb01a}, 
and parton cascade models \cite{zhang99a,molnar02a,lin02b}.
In this paper, we will study the elliptic flow using a parton
cascade model. We will first introduce the elliptic flow, and the
parton cascade model used for this study. Then, we will use the parton
cascade model to study the elliptic flow produced from two
different initial transverse distributions, one proportional to
the number of binary collisions, one proportional to the number
of wounded nucleons. We demonstrate that even though, elliptic
flow as a function of the impact parameter is very sensitive to
the initial transverse distribution of particles, elliptic flow
as a function of the relative multiplicity is almost independent of
the transverse distribution. Furthermore, we show that the minimum bias
differential elliptic flow is also insensitive to the initial
transverse particle distribution.

Elliptic flow is the elliptic deformation in the particle
transverse momentum distribution \cite{ollitrault92a}. 
It is usually characterized by
the second Fourier coefficient of the particle azimuthal 
distribution \cite{voloshin96a}. 
If we use $f(\phi)$ for the azimuthal distribution,
and choose the azimuthal angle of the reaction plane to
be zero, then
\begin{equation}
f(\phi)=v_0+2 v_1\cos(\phi)+2 v_2\cos(2\phi)+\cdots .
\end{equation}
The coefficient, $v_2$, is the elliptic flow observable. It is the
average of $\cos(2\phi)$ of produced particles. If the transverse
components of the momenta are know, it can be calculated by:
\begin{equation}
v_2=\left\langle\frac{p_x^2-p_y^2}{p_x^2+p_y^2}\right\rangle.
\end{equation}
In the above formula, $\langle\cdots\rangle$ 
denotes the average over particles.
Since initial momentum
distribution is isotropic, the elliptic flow is generated by final
state interactions. Final state interactions (or pressure gradient)
turn(s) the spatial anisotropy into momentum space anisotropy. It
has been shown that the elliptic flow is very sensitive to the
initial stage evolution, and can be used as a sensitive probe of
early dynamics \cite{sorge97a,sorge99a}.

In the following, we are going to use Zhang's Parton Cascade (ZPC) 
\cite{zhang98a} to
study elliptic flow at relativistic energies. The initial conditions
are set up similar to those in a recent study by Molnar's Parton
Cascade (MPC) \cite{molnar02a}. In
the local rest frame, the initial momentum distribution is thermal,
with a temperature of 700 MeV. Particles are uniformly distributed
between a space time rapidity of -5 and +5. The particle formation
proper time is 0.1 fm/c. There are totally 2100
gluons per central event. As the momentum transport is determined by 
the momentum opacity, the following results are also correct if the
total number of particles increases and the transport cross section
decreases by the same factor. To efficiently simulate momentum transport,
we use isotropic differential cross sections that preserve the
reaction plane of a collision. The total parton-parton elastic cross 
section $\sigma_{gg}$ will be varied to have values of 40 mb, 20 mb, 
10 mb to 
study the response of the system. These cross sections are effective 
cross sections as no radiative energy loss or parton recombinations
are included in the calculations.

We will study the elliptic flow
produced from two different initial transverse spatial distributions.
One is proportional to the number of binary collisions per unit area.
In this case, the particle number as a function of the impact parameter
is also proportional to the number of binary collisions.  The other
distribution is proportional to the number of wounded nucleons 
per unit area and the particle number as a function of the impact 
parameter is also
proportional to the number of wounded nucleons. In generating the
above distributions, the nucleon-nucleon inelastic cross section
$\sigma_{NN}$ is taken to be 40 mb, and the three parameter 
Woods-Saxon distribution
is used for the nucleons inside one nucleus. These two spatial
distributions are related to hard and soft particle production
mechanisms, respectively \cite{kolb01a}. 

The above initial spatial distribution and initial momentum
distribution factorize. As local densities are sampled for the
evolution of the expanding parton system according to the
Boltzmann equation, the factorization is not automatically conserved.
We also note that a geometry
with sharp cylindrical nuclei always leads to larger elliptic flow
values compared to the binary collision scaling case 
\cite{gyulassy01a,zhang99a,molnar02a,molnar02b,rak03a}.

\begin{figure}[htb]
\begin{center}
\includegraphics[width=3.2in,height=3.2in,angle=0]{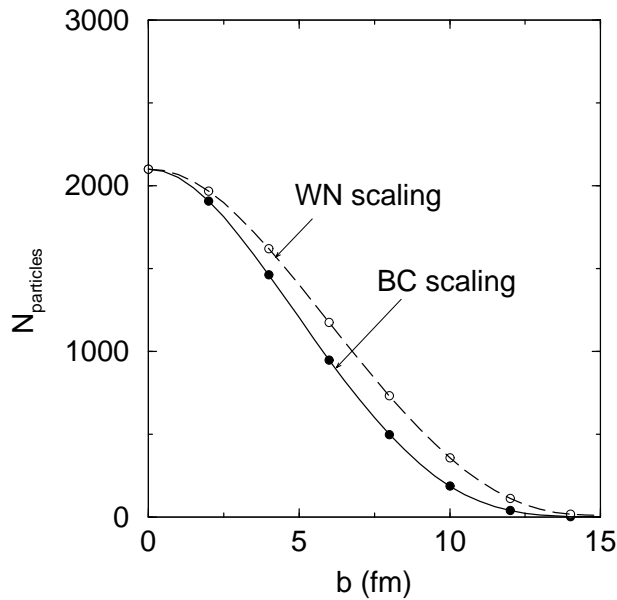}
\end{center}
%\null\vspace{1cm}
%\centerline{\epsfig{file=dndy3.eps,width=4.2in,height=4.2in,angle=-90}}
%\centerline{\epsfig{file=dndy3.eps,width=2.8in,height=2.8in,angle=-90}}
%\null\vspace{1cm}
\caption{
Number of particles as a function of the impact parameter for the
binary collision (BC) scaling case and the wounded nucleon (WN) 
scaling case. Circles are generated by the simulation code.
}
\label{npart_b}
\end{figure}

\begin{figure}[htb]
\begin{center}
\includegraphics[width=3.2in,height=3.2in,angle=0]{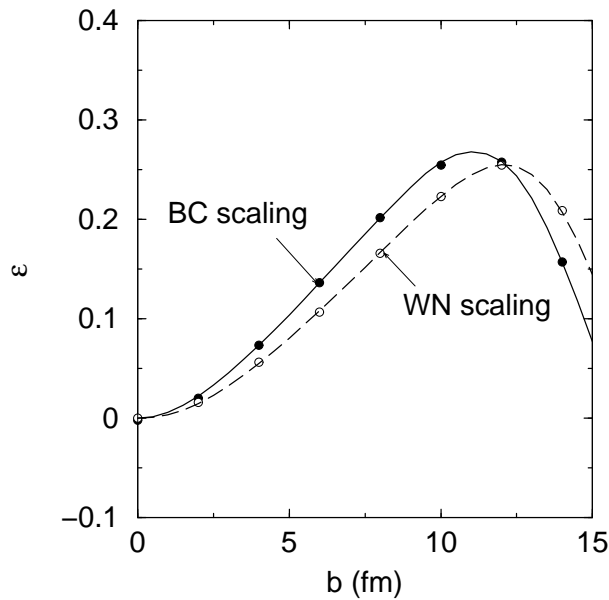}
\end{center}
\caption{
Initial spatial ellipticity as a function of the impact parameter.
}
\label{ell_b}
\end{figure}

Fig.~\ref{npart_b} gives the number of particles as a function of the impact
parameter. While the two distributions have the same
number of particles when $b=0$, the wounded nucleon scaling
has more particles than the binary collision scaling case. In
particular, when $b=10$ fm, the wounded nucleon scaling has about
twice as many particles as the binary collision scaling case.
Fig.~\ref{ell_b} has the initial spatial ellipticity as a function of
the impact parameter. The initial spatial ellipticity in the figure
is defined through
\begin{equation}
\epsilon=\left<\frac{y^2-x^2}{y^2+x^2}\right>.
\end{equation}
Note that this definition calculates the ratio first and then
the average and the magnitude is smaller than the ratio of
the averages. 

We first study the impact parameter dependence of the elliptic flow.
The set up is similar to the recent MPC model study \cite{molnar02a}. We 
approach the Boltzmann limit by increasing the number of particles
and at the same time decreasing the cross section by the same
factor \cite{zhang99a,molnar02a,zhang98b,cheng02a}. 
In the binary collision scaling case, the rescaling factors for 
$b=0,2,4,6,8,10,12,14$ fm are 
$\lambda=100,100,100,220,450,1100,5000,50000$. In the wounded nucleon
scaling case, the rescaling factors are
$\lambda=100,100,100,200,300,600,2000,11000$. The convergence is checked
by comparing to calculations done with $\lambda/2$. The $v_2$ is
calculated for particles with a rapidity range of $|y|<2$. 

\begin{figure}[htb]
\begin{center}
\includegraphics[width=3.2in,height=3.2in,angle=0]{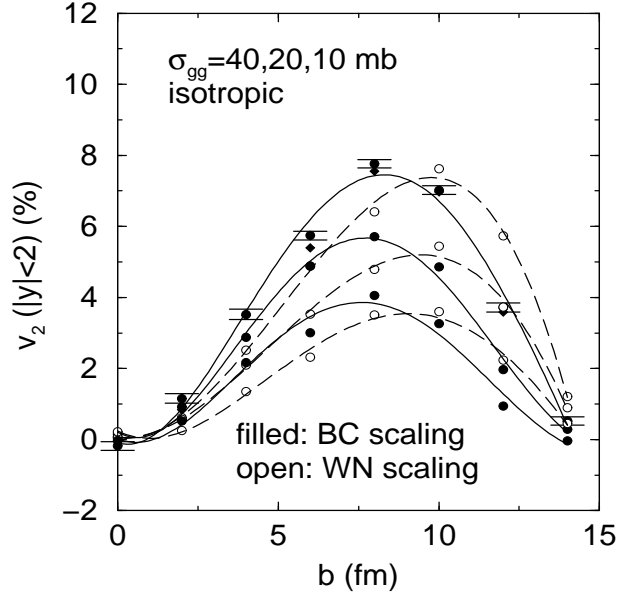}
\end{center}
\caption{
Elliptic flow as a function of the impact parameter. Filled
symbols are for the binary collision scaling case and open
symbols are for the wounded nucleon scaling case. The curves
are used to guide the eyes. Going from above, the three
sets of results are for $\sigma_{gg}=40,20,10$ mb, respectively.
For the binary collision scaling with $\sigma_{gg}=40$ mb case,
the statistical error bars are also drawn. They are about
the same for other curves. The diamonds are results for
the binary collision case with $\sigma_{gg}=40$ mb and parton
number rescaling factor of $\lambda/2$. They agree well
with the case with $\lambda$ particle division.
}
\label{v2_b}
\end{figure}

From Fig.~\ref{v2_b}, we
observe that as the total cross section increases, or more precisely,
as the transport cross section increases, the elliptic flow increases.
The binary collision scaling case is larger than the wounded nucleon
scaling case for small impact parameters and smaller than the wounded
nucleon scaling case for large impact parameters. This follows
the trend of the initial spatial ellipticity. 
However, the $v_2$ curves peak
at smaller impact parameters than the $\epsilon$ curves. This
indicates that both initial ellipticity and initial particle
density play roles in determining the elliptic flow. As the impact
parameter increases, the elliptic flow increases with the initial
ellipticity, however, after a point, the particle density is
not high enough to generate enough response and the elliptic flow
can not catch up with the initial ellipticity. It starts decreasing
with increasing impact parameter.

\begin{figure}[htb]
\begin{center}
\includegraphics[width=3.2in,height=3.2in,angle=0]{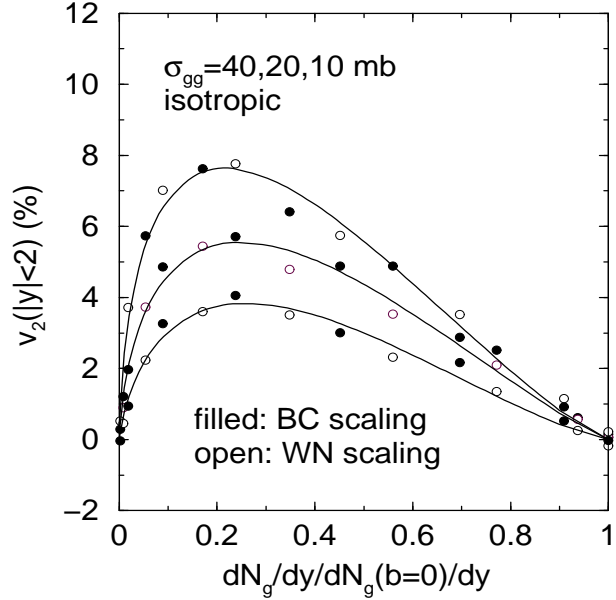}
\end{center}
\caption{
Elliptic flow as a function of the relative rapidity density.
Meanings of symbols are the same as those in Fig.~\ref{v2_b}.
}
\label{v2_n}
\end{figure}

An alternative way of characterizing centrality is to use
the relative central rapidity density, which is the ratio
of the central rapidity density to that in central collisions
with $b=0$.
If we plot the elliptic flow as a function of the relative
central rapidity density as in Fig.~\ref{v2_n}, we see that
for the same transport cross section, the two curves
corresponding to the binary collision scaling and the
wounded nucleon scaling overlap. In other words, the impact
parameter dependence of elliptic flow is cancelled by
the impact parameter dependence of the multiplicity.
Similar observations have
been made in a recent hydrodynamics study \cite{kolb01a}. 
This indicates
that if we use the relative central rapidity density as
a measure of centrality, the elliptic flow is not 
sensitive to whether the initial distribution is binary
collision scaling or wounded nucleon scaling. It reflects
the particle transport cross section, or in the case of
hydrodynamics, the equation of state. Because
of the viscous corrections, the cascade calculations have
a bend over when the relative central rapidity density
is small. In contrast, the hydrodynamic calculations
have an almost straight line dependence and overshoot data when
the relative rapidity density is small. 
We also note that the $\sigma_{gg}=40$ mb
binary collision scaling case is consistent with set D of
\cite{molnar02a}.

\begin{figure}[htb]
\begin{center}
\includegraphics[width=3.2in,height=3.2in,angle=0]{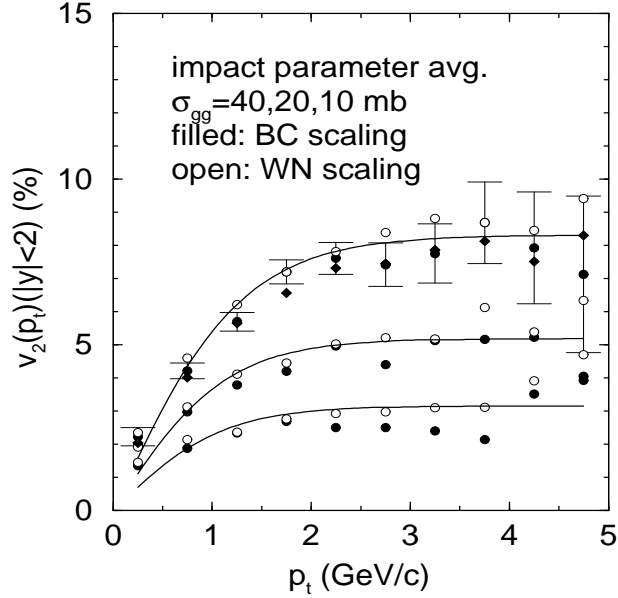}
\end{center}
\caption{
Impact parameter averaged differential elliptic flow as a function of 
the transverse momentum. Meanings of symbols are the same as those
in Fig.~\ref{v2_b}.
}
\label{v2pt_min}
\end{figure}

Now we turn to the study of $p_t$ differential elliptic flow which
can give further information about the evolution \cite{li99a}. 
Hydrodynamic
studies agree well with low $p_t$ data. At $p_t>2 $ GeV, the data
are consistent with a constant behavior while hydrodynamic results
keep on increasing. In dynamic models, only when viscous effects
are taken into account, is it possible to describe the deviation
from the ideal hydrodynamical behavior.  A recent hydrodynamic study
demonstrates that the minimum bias $p_t$ differential flow
is not sensitive to the initial nuclear profile. We want to know whether
it is also true when viscous effects are taken into account. 
Fig.~\ref{v2pt_min} shows the impact parameter averaged $p_t$ 
differential flow. 
The calculations with binary collision scaling agree well with
those with wounded nucleon scaling. This is true not only
for the low $p_t$ region, but also for the high $p_t$ region where
viscous effects are important. 

\begin{figure}[htb]
\begin{center}
\includegraphics[width=3.2in,height=3.2in,angle=0]{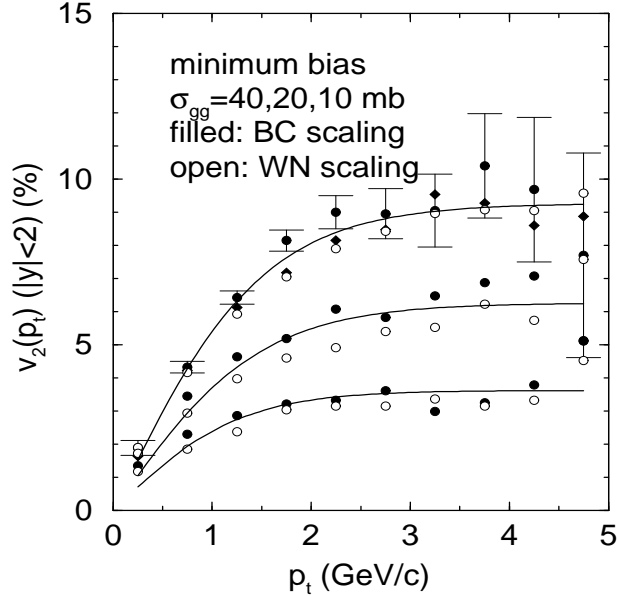}
\end{center}
\caption{
Minimum bias differential elliptic flow as a 
function of the transverse momentum. Meanings of symbols are the
same as those in Fig.~\ref{v2_b}.
}
\label{v2pt_min_star}
\end{figure}

Another way of averaging over events is to calculate the multiplicity
weighted average of cosine of the azimuthal angle. This gives the 
minimum bias elliptic flow. Results from the ZPC model are shown in 
Fig.~\ref{v2pt_min_star}. The binary collision scaling
and the wounded nucleon scaling agree well with each other. This further
demonstrates that the minimum bias $p_t$ differential elliptic flow
is insensitive to the initial nuclear profile. A comparison of
Fig.~\ref{v2pt_min} and Fig.~\ref{v2pt_min_star} shows that
the minimum bias results are about 10\% higher than
the impact parameter averages.  
Hence the impact parameter averaged differential elliptic flow
can be considered as a reasonably good approximation of the 
minimum bias differential elliptic flow. As pointed out 
in \cite{molnar02a}, the minimum bias elliptic flow
weights in more central events. The central events in the 
wounded nucleon scaling case have lower elliptic flow than those
in the binary collision scaling case. This can lead to a relatively
smaller minimum bias elliptic flow in the wounded nucleon scaling case.
However, the decrease in the relative amplitude
can not be determined with the current statistics.

In summary, we demonstrate that the elliptic flow as a function
of relative central rapidity density  in the binary scaling 
case agrees well with that in the wounded nucleon case.
In addition, the minimum bias
$p_t$ differential elliptic flow in the binary scaling case
also agrees well with that in the wounded nucleon case. This
is true not only for the low $p_t$ region, but also for the
high $p_t$ region where viscous effects are important and
ideal hydrodynamics deviates from experimental data. As hadronization
will not change the high $p_t$ elliptic flow \cite{molnar02a},
and recent research indicates the possibility of extracting
the parton elliptic flow from hadron elliptic flow \cite{lin03a},
the elliptic flow as a function of relative rapidity density
and the minimum bias $p_t$ differential elliptic flow
are promising observables for the extraction of information
about final state interactions \cite{voloshin00a}.

This work was supported by the U.S. National Science 
Foundation under Grant No. 0140046. 
We thank P. Kolb, Z.W. Lin, D. Molnar, and A. Poskanzer for helpful
discussions. We also thank the Parallel Distributed
System Facility at the National Energy Research Scientific Computer Center
for providing computer resources.

% \appendix

% \section{}
% \label{}

\end{document}